\documentclass[twocolumn,prl,aps]{revtex4}
\usepackage{amsmath}
\usepackage{amssymb}
\usepackage{amsfonts}
\usepackage{graphicx}

\begin{document}
\title
{Dispersion of Volume Relativistic Magnetoplasma Excitation in a Gated Two-Dimensional Electron System}

\author{P.~A.~Gusikhin}
\email{gusikhin@issp.ac.ru}
\author{V.~M.~Muravev}
\author{I.~V.~Kukushkin}
\affiliation{Institute of Solid State Physics, RAS, Chernogolovka, 142432 Russia}

\date{\today}

\begin{abstract}
The dispersion of the volume relativistic magnetoplasma mode in a gated GaAs/AlGaAs quantum well is measured using a coupled resonators detection technique. The weakly damped relativistic mode exhibits an unusual zigzag-shaped magnetodispersion dependence dictated by the diagonal component of the resistivity tensor $\rho_{xx}$. The plasma excitation easily hybridizes with photon modes due to a large spatial delocalization of its electromagnetic field. The effects of electron density and structure geometry on the excitation spectrum have been investigated.
\end{abstract}


\maketitle

The last few decades have witnessed a large surge of research in the field of plasmonics~\cite{Bryant}. An important impetus comes from the physical property of plasma waves to confine electromagnetic radiation at subwavelength scales. Further, plasma waves may be efficiently manipulated on a chip. The plasmonics concept has already found a wealth of applications for devices working in the optical part of the spectrum~\cite{Barnes:03, Ozbay:06}. However, recent progress in the purity of semiconductor structures has made it possible to adapt the plasmonic phenomena from the optical region of the spectrum to the microwave and terahertz (THz) bands. The THz band has attracted significant attention in recent years, due to its scientific richness and promise of many exciting new applications. Potential applications span from non-destructive testing, security imaging, and fast telecommunication to biomedical applications~\cite{Zhang}.

Standard two-dimensional (2D) plasmons are observable at frequencies of $\omega >1/\tau$, where the relaxation time $\tau(T)$ essentially decreases with increasing temperature~\cite{Andreev}. This presents a challenge because the plasmonic effects are visible only at sufficiently large frequencies and sufficiently low temperatures. Indeed, 2D plasma waves at frequencies suitable for applications with $f < 500$~GHz have been observed only at temperatures $T<80$~K~\cite{Knap, Eisentstein, Shaner:05, Knap:08, Muravjov:10, Muravev:APL}. One way to circumvent this imposed limitation is to deal with relativistic plasma excitations. These unusual weakly damped plasma waves are excited in high-conductivity ($2 \pi \sigma > c$ in Gaussian units) gated two-dimensional electron systems (2DESs)~\cite{Muravev:15, Gusikhin:14}. It has been proven that such relativistic plasmons survive at temperatures up to $300$~K in the microwave frequency range. Despite the fact that plasma excitations in high conductivity 2DESs should become more photon-like, and lose sensitivity to the density of two-dimensional electrons and magnetic-field magnitude~-- frequency of the relativistic plasma mode retains a strong electron density and external magnetic field dependence even in the extreme $2 \pi \sigma/c \gg 1$ regime. 

\begin{figure}[!t]
\includegraphics[width=1\linewidth]{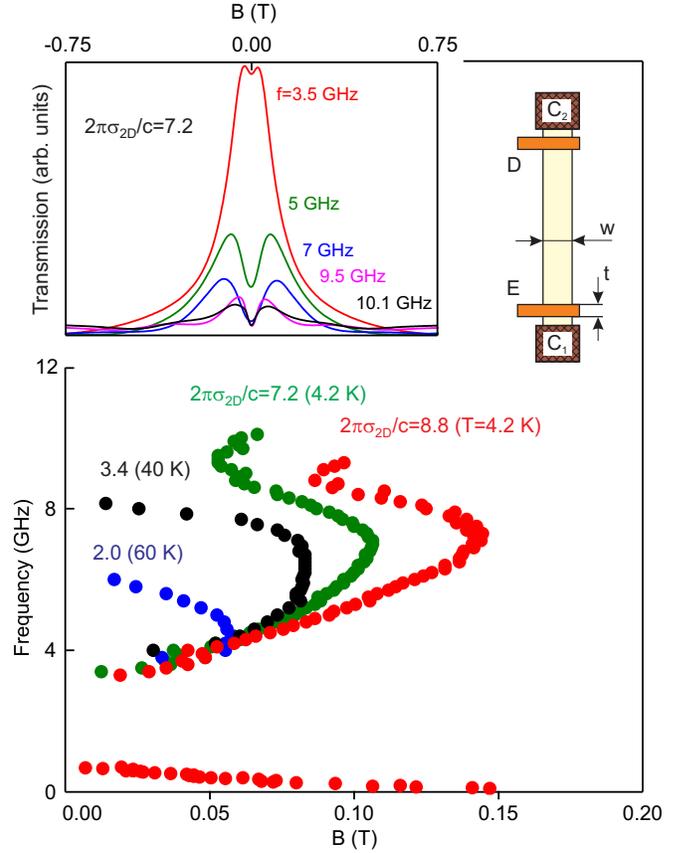}
\caption{The B-field dependencies of the magnetoplasma resonance frequencies in a sample with $w=50$~$\mu$m, $t=40$~$\mu$m and an electron density of $1.8\times 10^{11}/{\rm cm}^2$ measured at different temperatures. The inset shows B-field dependencies of the microwave transmission measured for several microwave frequencies. In the upper right corner, a schematic view of the sample is shown. It is a 2DES stripe with ohmic contacts on both ends and two symmetric gates near the contacts.} 
\label{1}
\end{figure}

The first studies of relativistic plasmons have concentrated mainly on the physical properties of the low-frequency edge mode~\cite{Muravev:15, Gusikhin:14}. However, understanding the high-frequency volume mode, which is more attractive for applications, is still missing. In this letter, we report on detailed experimental research of the volume relativistic plasma excitation. We found that volume and edge relativistic plasma excitations exhibit drastically different behavior for most 2DES parameters. This fact indicates that the electromagnetic fields of these two modes have different spatial symmetry.  

The experimental results presented below were obtained on a 2DES fabricated from modulation-doped GaAs/AlGaAs heterostructures. We used two structures with electron densities of $n_s=1.8\times 10^{11}$~cm$^{-2}$ ($2 \pi \sigma/c = 7.2$ and $2 \pi \sigma/c = 8.8$ at $T=4.2$~K) and one structure with electron density of $4.3\times 10^{11}$~cm$^{-2}$ ($2 \pi \sigma/c = 5.5$ at $T=4.2$~K). The distance between the 2DES and the crystal surface for all structures was $h=190$~nm. Stripe-shaped mesas were fabricated from these heterostructures with widths $w=30$~$\mu$m, $50$~$\mu$m, and $100$~$\mu$m (Fig.~1). $680$~$\mu$m long mesas were ended in grounded ohmic contacts, $\rm C_1$ and $\rm C_2$, at both ends. At a distance of $40$~$\mu$m from these contacts two identical metallic gates were deposited on top of the sample across the mesa. The widths of the gates were $t=10$~$\mu$m and $40$~$\mu$m. Sample with dimensions $3\times 1\times 0.65$~mm$^3$ was mounted on the ceramic chip-carrier. Chip-carrier has two SMA connectors and impedance-matched coplanar waveguides for the signal transmission between connector and sample. Connection between coplanar waveguide and the sample was carried out using aluminum wires with diameter $50$~$\mu$m and ultrasonic bonding. Microwave radiation with frequency varying from $0.1$ to $20$~GHz was modulated at $2$~kHz and guided through the coaxial cable into the cryostat. Through the coplanar waveguide, the radiation was supplied to the excitation gate, $\rm E$. In our previous studies, we showed that parts of the 2DES stripe between gate $\rm E$ and contact $\rm C_1$ and, correspondingly, between gate $\rm D$ and contact $\rm C_2$ form an identical plasmonic resonators for the relativistic mode under investigation~\cite{Muravev:15, Gusikhin:14}. A microwave signal supplied to gate $\rm E$ excites plasma modes both in the native $\rm E$--$C_1$ plasmonic resonator, and via the induced $ac$ voltage, in the remote $\rm D$--$C_2$ resonator. The transmitted resonant signal was detected synchronously from gate $\rm D$ with the help of a Shottky diode placed outside the cryostat. The experiments were carried out at a sample temperature in the range of $T=4.2$--$77$~K.

\begin{figure}[!t]
\includegraphics[width=1\linewidth]{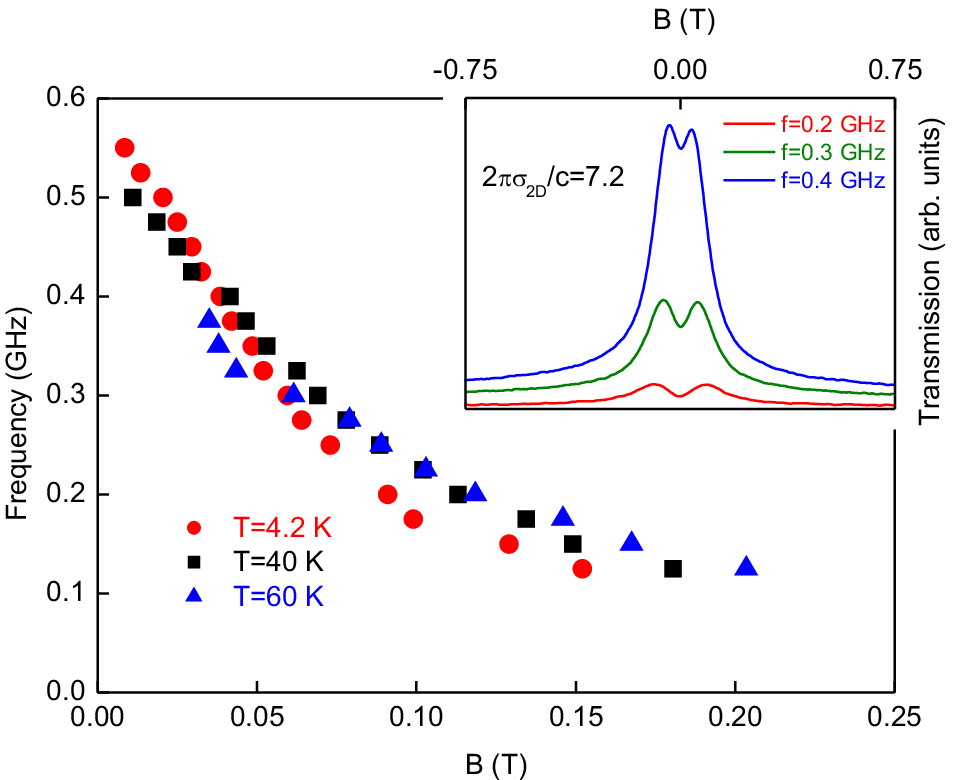}
\caption{Magnetodispersion of the lower branch in a sample with $w=50$~$\mu$m, $t=40$~$\mu$m and an electron density of $1.8\times 10^{11}/{\rm cm}^2$ measured at different temperatures. Inset shows B-field dependencies of the microwave transmission measured for several microwave frequencies at $T=4.2$~K.} 
\label{2}
\end{figure}

In Fig.~1, the $B$-field dependencies of the magnetoplasma resonance frequencies are plotted for a samples with gates dimensions $w=50$~$\mu$m and $t=40$~$\mu$m ($n_s=1.8\times 10^{11}$~cm$^{-2}$). The experiments were conducted at three different temperatures, $T=4.2$, $40$, and $60$~K. The corresponding values of conductivities ($2\pi\sigma/c$) measured at $B=0$~T by a four-probe transport are listed in the same figure. The magnetodispersion has two branches separated by a frequency gap. The low-frequency branch with a negative magnetodispersion is reminiscent of edge magnetoplasma mode (EMP)~\cite{Allen:83, Volkov:88, Shikin}. The excitation frequency drops with increasing $B$ as $\omega \sim \sigma_{xy}q \propto  n_s q/B$ ($q$ being an EMP wavevector), and it is confined to the edge in strong magnetic fields. The high-frequency branch has a positive magnetodispersion analogous to the volume magnetoplasma mode. The branch exhibits zigzag-shaped behavior at higher frequencies. Such behavior indicates a large polaritonic contribution to the mode formation~\cite{Kukushkin:03, Muravev:13}. 

\begin{figure}[!t]
\includegraphics[width=1\linewidth]{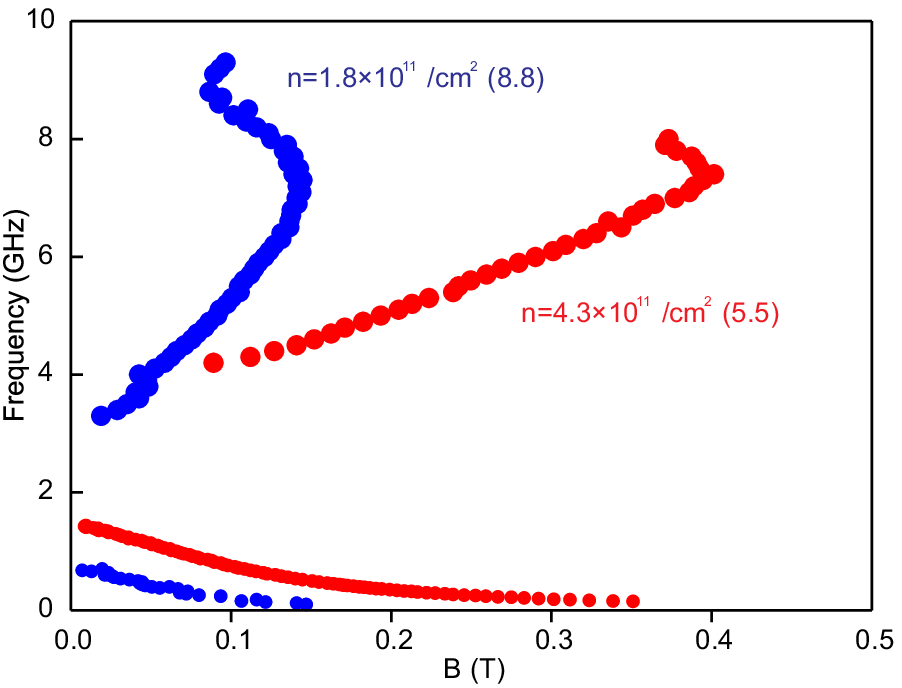}
\caption{Magnetodispersion of the lower branch in a sample with $w=50$~$\mu$m, $t=40$~$\mu$m and an electron density of $1.8\times 10^{11}/{\rm cm}^2$ measured at different temperatures. Inset shows B-field dependencies of the microwave transmission measured for several microwave frequencies at $T=4.2$~K.} 
\label{3}
\end{figure}

Figure~2 shows a zoomed magnetodispersion of the low-frequency edge mode measured for the same set of temperatures. The most important observation is that unlike ordinary 2D plasmons, both edge and volume relativistic modes have a strong dependence on 2D conductivity, $\sigma$, and 
disappear at $2 \pi \sigma/c \approx 1$ (Gaussian units are used in the present Letter unless
otherwise stated)~\cite{Muravev:15, Gusikhin:14}. However, from Fig.~1 it is obvious that the temperature increase from $4.2$~K to $60$~K has a much more pronounced impact on the volume relativistic magnetoplasma excitation than that on the edge excitation. We attribute this observation to the fact that behavior of the relativistic modes in the magnetic field is determined by different components of the 2D conductivity tensor $ \hat{\sigma}$. Indeed, in the limit of $\omega_c \gg \omega$, where $\omega_c = eB/m^{\ast}c$ is the cyclotron frequency, the tensor $\hat{\sigma}$ is described by the following expression:    
\begin{equation}
\hat{\sigma} = \left ( \begin{matrix} \sigma_{xx} & \sigma_{xy} \\  \sigma_{yx} & \sigma_{yy} \end{matrix} \right )  = \frac{n_s e^2 \tau}{m^{*}\left(1 + (\omega_c \tau)^2\right)} \left ( \begin{matrix} 1 & -\omega_c \tau \\  \omega_c \tau & 1 \end{matrix} \right ).
\end{equation}
Here, $\tau$ is transport relaxation time. In the limit of strong magnetic field, dynamics of the edge relativistic magnetoplasma mode is governed by the temperature-independent parameter, that is determined by the conductivity as:
\begin{equation}
\frac{\sigma_{xx}^2+\sigma_{xy}^2}{\sigma_{xy}}=\frac{1}{\rho_{xy}}=\frac{ne^2}{m^{*}\omega_c}.
\end{equation}
We believe that magnetodispersion of the highly temperature-sensitive volume relativistic mode is in turn intimately connected to the conductivity as:
\begin{equation}
\frac{\sigma_{xx}^2+\sigma_{xy}^2}{\sigma_{xx}}=\frac{1}{\rho_{xx}}=\frac{ne^2\tau}{m^{*}}.
\end{equation}

\begin{figure}[!t]
\includegraphics[width=1\linewidth]{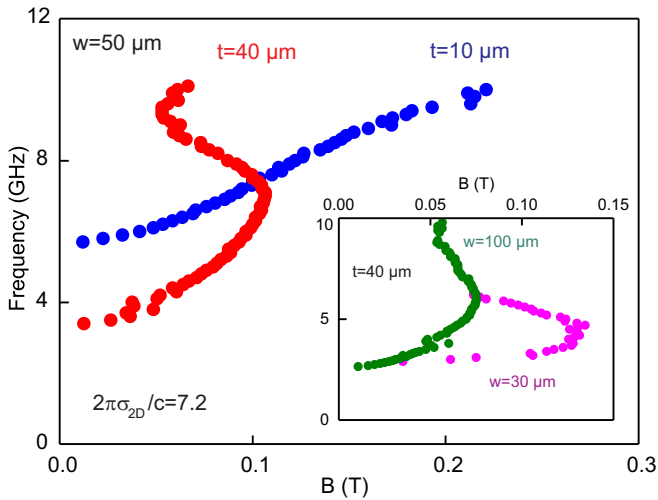}
\caption{Magnetic field dependencies of the plasma mode for samples with equal 2DES widths $w=50$~$\mu$m and different gate widths: $t=10$~$\mu$m and $t=40$~$\mu$m. The inset shows the Magnetic field dependencies of plasma mode for samples with equal gate widths $t=40$~$\mu$m and different 2DES widths: $w=30$~$\mu$m and $w=100$~$\mu$m.} 
\label{4}
\end{figure}

Figure~3 shows the magnetodispersion curves of the relativistic plasma modes obtained for samples with electron densities $n_s=1.8\times 10^{11}$~cm$^{-2}$ and $4.3\times 10^{11}$~cm$^{-2}$. The most interesting observation is that the functional magnetic field dependency of the volume mode frequency undergoes a significant change with increasing electron density. The mode becomes more photon-like at higher electron density. This observation, as well as the zigzag-like shape of the magnetodispersion curves, can be accounted for by a hybridization of the volume plasmon with a photon mode~\cite{Muravev:11, Muravev:13}. Indeed, the branch originates at $B = 0$~T from a certain fundamental plasma frequency, $\omega_p (1)$, and tends to a photonic horizontal asymptotic value, $\omega_0$. However, at some magnetic field value, the mode exhibits a return jump to the magnetodispersion of the second plasmon harmonics, $\omega_p (2)$, which already mixes with the second photonic harmonics, $2 \omega_0$. We think that the photon mode originates from the resonance of the electromagnetic wave within the body of the semiconductor chip under study. From the geometrical sizes of the chip, we deduce $f_0=\omega_0/(2 \pi) = 13$~GHz. With decreasing two-dimensional electron density, the plasmon contribution to the hybrid polaritonic excitation becomes smaller, and hence the magnetodispersion appears closer to the photon dispersion line in the moderate magnetic fields~\cite{Muravev:11, Muravev:13}. It should be noted that both edge and volume relativistic modes demonstrate square-root dependence between plasma frequency and electron density $\omega_p \sim \sqrt{n_s}$ at $B=0$~T, that is common for plasmons.

In order to determine the influence of the geometrical parameters on the volume relativistic mode properties, we conducted several additional experiments. Earlier it was experimentally shown that the zero-field frequency of the edge relativistic plasma excitation obeys the relationship~\cite{Muravev:15, Gusikhin:14}
\begin{equation}
\omega_p\sim\frac{1}{\sqrt{w\times t}},
\label{disp}
\end{equation}
where $w$ and $t$ are the widths of the 2DES stripe and the gate respectively (see Fig.~1). For the volume mode we see a different behavior. Fig.~4 shows the magnetodispersion of the high-frequency volume mode for samples with different gate sizes. A zero-field frequency of the mode shows clear dependence on the gate width $t$. Two points we obtained are fit to typical square root law. However, a substantial change in the 2DES strip width from $w=30$~$\mu$m to $100$~$\mu$m leads to virtually no change in the zero-field frequency. We attribute such a difference between the volume and edge relativistic modes to different spatial symmetry of their electromagnetic fields. 

\begin{figure}[!t]
\includegraphics[width=1\linewidth]{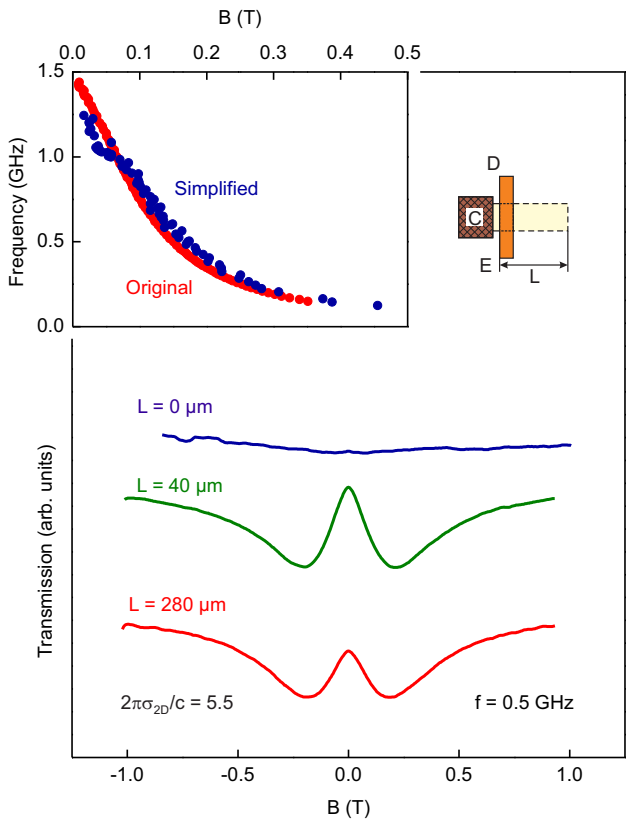}
\caption{Magnetic field dependencies of microwave transmission through the gate for the samples with different distance between the end of 2DES stripe and the gate, $L=240$~$\mu$m, $L=0$~$\mu$m, and for the sample without 2DES under gate at the frequency $f=0.5$~GHz. Inset shows the magnetodispersions of the edge magnetoplasma excitations obtained for the original and simplified geometries with the same dimensions $w=50$~$\mu$m and $t=40$~$\mu$m. In the upper right corner, a schematic view of the samples is shown.} 
\label{5}
\end{figure}

Determining where the plasma mode is localized is of particular importance to the research and applications.  Additional experiments were conducted in order to ascertain this information. Several samples with simplified geometry (Fig.~5) were fabricated from the structure with electron density of $4.3\times 10^{11}$~cm$^{-2}$ ($2 \pi \sigma/c = 5.5$ at $T=4.2$~K). This geometry was similar to the original one (Fig.~1), except that there was only one contact and gate.  The dimensions in the series of samples were identical~-- $w=50$~$\mu$m and $t=40$~$\mu$m~-- aside from $L$, the distance between the end of the 2DES stripe and the gate (Fig.~5). This distance amounted to $L=0$~$\mu$m (no 2DES under the gate), $40$~$\mu$m, and $280$~$\mu$m in different samples. We measured microwave transmission through the gate from its $\rm E$ side to the $\rm D$ end. Figure~5 shows the magnetic field dependencies of the microwave transmission measured for different samples on the same frequency $f=0.5$~GHz. For the curves corresponding to $L=280$~$\mu$m and $L=40$~$\mu$m, there are symmetrical absorption resonances indicating relativistic plasmon excitation. Resonances on these two curves match in position, amplitude, and width. Moreover, magnetodispersion curves obtained for the samples with modified geometry coincide with the magnetodispersion curve for the original geometry (inset to Fig.~5). In contrast, there are no resonances on the magnetic field dependency of transmission obtained for the sample without 2DES under the gate. These results suggest that the mode is localized between the contact and the gate, including the gated area of 2DES. 

In conclusion, we have investigated the high-frequency volume relativistic plasma mode in a gated GaAs/AlGaAs quantum well. This mode exhibits an unusual zigzag-shaped magnetodispersion dependence due to its strongly polaritonic nature. It was found that the volume and edge relativistic plasma excitations have completely different behavior with respect to most 2DES parameters. These shows that the electromagnetic fields of this two modes have different symmetry. We have determined that in our geometry plasma excitation is localized between the contact and the gated region of 2DES. The differences in temperature dependence between two modes led us to conclude that the dynamics of the edge relativistic mode is governed by the temperature-insensitive $\rho_{xy}$ component of the 2DES resistivity tensor, while the volume relativistic mode is connected to $\rho_{xx}$. Nonetheless volume mode exists at relatively high temperature, allowing it to be attractive for applications in plasmonic devices.

The authors gratefully acknowledge financial support from the Russian Scientific Fund (Grant No.~12-02-00590) and President Grant (Grant No. MK-5549.2015.2).


\begin{thebibliography}{99}

\bibitem{Bryant}
M.~Pelton and G.~Bryant, \textit{Introduction to Metal-Nanoparticle Plasmonics} (Wiley, New Jersey, 2013).

\bibitem{Barnes:03}
W.~L.~Barnes, A.~Dereux, T.~W.~Ebbesen, Nature {\bfseries{424}}, 824 (2003).

\bibitem{Ozbay:06}
E.~Ozbay, Science {\bfseries{311}}, 189 (2006).

\bibitem{Zhang}
X.-C.~Zhang and Jingzhou~Xu, \textit{Introduction to THz Wave Photonics} (Springer, London, 2010). 

\bibitem{Andreev}
I.~V.~Andreev, V.~M.~Muravev, V.~N.~Belyanin, and I.~V.~Kukushkin, Appl.~Phys.~Lett. {\bfseries{105}}, 202106 (2014).

\bibitem{Knap}
W.~Knap, Y.~Deng, S.~Rumyantsev, and M.~S.~Shur, Appl.~Phys.~Lett. {\bfseries{81}}, 4637 (2002).

\bibitem{Eisentstein}
X.~G.~Peralta, S.~J.~Allen, M.~C.~Wanke, N.~E.~Harff, J.~A.~Simmons, M.~P.~Lilly, J.~L.~Reno,
P.~J.~Burke, and J.~P.~Eisenstein, Appl.~Phys.~Lett. {\bfseries{81}}, 1627 (2002).

\bibitem{Shaner:05}
E.~A.~Shaner, Mark~Lee, M.~C.~Wanke, A.~D.~Grine, J.~L.~Reno, and S.~J.~Allen, Appl.~Phys.~Lett. {\bfseries{87}}, 193507 (2005).

\bibitem{Knap:08}
W.~Knap, F.~Teppe, N.~Dyakonova, D.~Coquillat, and J.~Lusakowski, J. Phys. Condens. Matter {\bfseries{20}}, 384205 (2008).

\bibitem{Muravjov:10}
A.~V.~Muravjov, D.~B.~Veksler, V.~V.~Popov, O.~V.~Polischuk, N.~Pala, X.~Hu, R.~Gaska, H.~Saxena, R.~E.~Peale, and M.~S.~Shur, Appl. Phys. Lett. {\bfseries{96}}, 042105 (2010).

\bibitem{Muravev:APL}
V.~M.~Muravev, I.~V.~Kukushkin, Appl.~Phys.~Lett. {\bfseries{100}}, 082102 (2012).

\bibitem{Muravev:15}
V.~M.~Muravev, P.~A.~Gusikhin, I.~V.~Andreev, and I.~V.~Kukushkin,
Phys.~Rev.~Lett. {\bfseries{114}}, 106805 (2015).

\bibitem{Gusikhin:14}
P.~A.~Gusikhin, V.~M.~Muravev, and I.~V.~Kukushkin, Pis'ma v Zh. Eksp. Teor. Fiz. {\bf 100},
732 (2014)[JETP Lett. {\bf 100}, 648 (2015)].

\bibitem{Allen:83} 
S.\,J. Allen, Jr., H.\,L. St\"{o}rmer, and
J.\,C.\,M. Hwang, Phys. Rev. B {\bf 28}, 4875 (1983).

\bibitem{Volkov:88}
V.\,A. Volkov and S.\,A. Mikhailov, Zh.~Eksp.~Teor.~Fiz. {\bf 94},
217 (1988)[Sov. Phys. JETP {\bf 67}, 1639 (1988)].

\bibitem{Shikin}
S.~S.~Nazin and V.~B.~Shikin, Zh.~Eksp.~Teor.~Fiz. {\bf 94},
133 (1988)

\bibitem{Kukushkin:03}
I.~V.~Kukushkin, J.~H.~Smet, S.~A.~Mikhailov, D.~V.~Kulakovskii, K.~von~Klitzing, and W.~Wegscheider,
Phys.~Rev.~Lett. {\bfseries{90}}, 156801 (2003).

\bibitem{Muravev:11}
V.~M.~Muravev, I.~V.~Andreev, I.~V.~Kukushkin, S.~Schmult, and W.~Dietsche, Phys.~Rev.~B {\bfseries{83}}, 075309 (2011).

\bibitem{Muravev:13}
V.~M.~Muravev, P.~A.~Gusikhin, I.~V.~Andreev, and I.~V.~Kukushkin, Phys.~Rev.~B {\bfseries{87}}, 045307 (2013).

\end{thebibliography}
\end{document}